\documentclass[twocolumn,aps,prl,showpacs,superscriptaddress,groupedaddress]{revtex4}
\usepackage[latin9]{inputenc}
\usepackage{color}
\usepackage{bm}
\usepackage{amsmath}
\usepackage{amssymb}
\usepackage{graphicx}
\usepackage{esint}

\makeatletter
\@ifundefined{textcolor}{}
{%
 \definecolor{BLACK}{gray}{0}
 \definecolor{WHITE}{gray}{1}
 \definecolor{RED}{rgb}{1,0,0}
 \definecolor{GREEN}{rgb}{0,1,0}
 \definecolor{BLUE}{rgb}{0,0,1}
 \definecolor{CYAN}{cmyk}{1,0,0,0}
 \definecolor{MAGENTA}{cmyk}{0,1,0,0}
 \definecolor{YELLOW}{cmyk}{0,0,1,0}
}

\usepackage{dcolumn}% needed for some tables
\usepackage{bm}% for math
\makeatother

\begin{document}
% the following line is for submission, including submission to the arXiv!!
%\hspace{5.2in} \mbox{Fermilab-Pub-04/xxx-E}

\title{No-pumping theorem for many particle stochastic pumps}

\author{Shahaf Asban}

\affiliation{Faculty of Physics ,Technion - Israel Institute of Technology, Haifa
32000, Israel}

\author{Saar Rahav}

\affiliation{Schulich Faculty of Chemistry ,Technion - Israel Institute of Technology,
Haifa 32000, Israel}
\begin{abstract}
Stochastic pumps are models of artificial molecular machines which
are driven by periodic time variation of parameters, such as site and
barrier energies. The no-pumping theorem states that no directed motion
is generated by variation of only site or barrier energies {[}S. Rahav,
J. Horowitz, and C. Jarzynski, Phys. Rev. Lett., \textbf{101 }, 140602
(2008){]}. We study stochastic pumps of several {\em interacting}
particles and demonstrate that the net current of particles satisfies
an additional no-pumping theorem.
\end{abstract}


\date{\today}

\maketitle
Molecular motors and machines are an essential component in living
organisms. They perform tasks such as carrying loads, contracting
muscles and many other crucial functions \cite{Howardbook}. Many
research groups are actively trying to venture beyond the motors found
in nature, aiming to design and synthesize artificial molecular machines
\cite{Kottas2005,Kay2007,Feringa2007,Michl2009,Coskun2012}.

Since artificial machines can be designed, it is possible to operate
them using new driving mechanisms which are not found in biological
motors and machines. One such driving mechanism is the rectification
of periodic time variation of external parameters. Due to the similarity
with everyday pumps such systems are often referred to as stochastic
pumps. Stochastic pumps are therefore closely related to thermal ratchets
\cite{Reimann2002}, but the term is more commonly used for systems
with a discrete set of coarse-grained states. The dynamics
of stochastic pumps have been investigated extensively in recent years
\cite{Parrondo1998,Sokolov1999,Astumian2003,Sinitsyn2007,Rahav2011,Chernyak2012},
see also the reviews by Sinitsyn \cite{Sinitsyn2009} and Astumian
\cite{Astumian2011} for an overview.

Motivated by a beautiful experiment on catananes \cite{Leigh2003},
and by work focused on an adiabatically driven model \cite{Astumian2007},
a no-pumping theorem (NPT) for stochastic pumps was found \cite{Rahav2008}.
This non adiabatic result identifies driving mechanisms which will
not lead to directed motion. In a parallel development, Chernyak and
Sinitsyn \cite{Chernyak2008} showed how to generalize the NPT to
account for the topology of the network of transitions between
states. Due to its simple structure, and somewhat non intuitive result,
the NPT has generated considerable interest \cite{Horowitz2009,Mandal2012,Maes2010,Mandal2011,Ren2011}.
All this body of work was focused on the stochastic dynamics of a
single particle. The aim of this letter is to investigate the validity
of the NPT for many particle stochastic pumps, where several {\em
interacting} particles jump between a set of binding sites. We demonstrate
that an NPT holds for this many particle system for any local interaction.
%Interestingly, this NPT is not obtained by a direct application of
%the NPT of \cite{Rahav2008} to the many particle pump.

\paragraph*{Single particle NPT-}

Consider a system which makes sudden transitions
between a set of coarse-grained states, labeled by $\alpha,\beta,\gamma,\cdots$
The transition are assumed to be Markovian and are characterized by
transition rates $R_{\beta\alpha}\ge0$ (for $\alpha\neq\beta$).
We assume that the system is connected, namely that it is possible
to reach any state from an arbitrary initial state in a finite number
of transitions. We furthermore assume that when $R_{\beta\alpha}>0$
then also $R_{\alpha\beta}>0$ (microreversibility).

The probability $p_{\alpha}(t)$ to find the system in state $\alpha$
evolves according to a master equation
\begin{equation}
\frac{d\bm{p}}{dt}=\bm{Rp},
\end{equation}
where the diagonal elements of $\bm{R}$ satisfy $R_{\alpha\alpha}=-\sum_{\sigma\neq\alpha}R_{\sigma\alpha}$,
thereby ensuring probability conservation.

\begin{figure}
\begin{centering}
\begin{minipage}[t]{1\columnwidth}%
\begin{center}
(a)\includegraphics[scale=0.16]{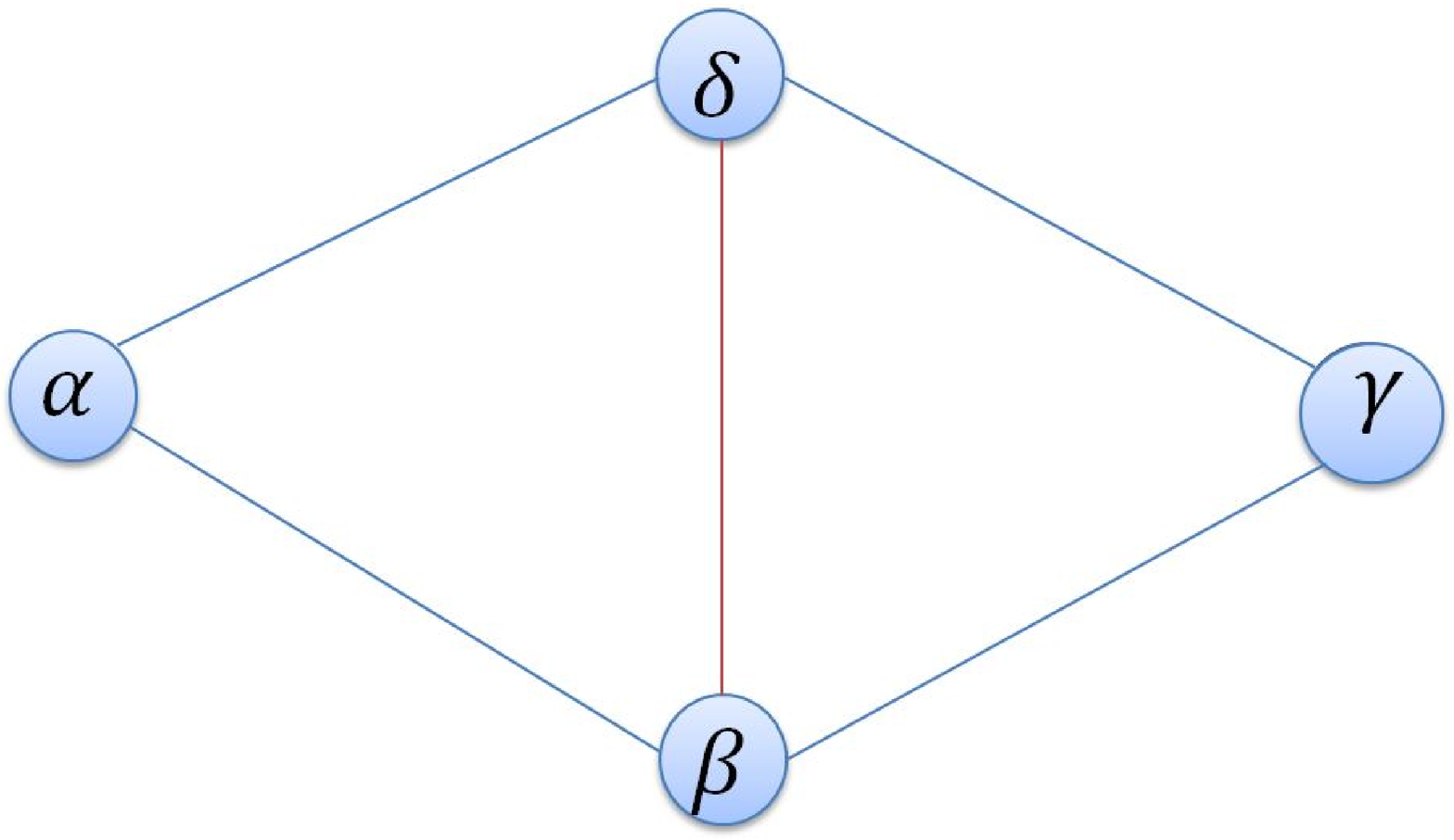}
\par\end{center}%
\end{minipage}
\par\end{centering}

\begin{centering}
\begin{minipage}[t]{1\columnwidth}%
\begin{center}
(b) \includegraphics[scale=0.17]{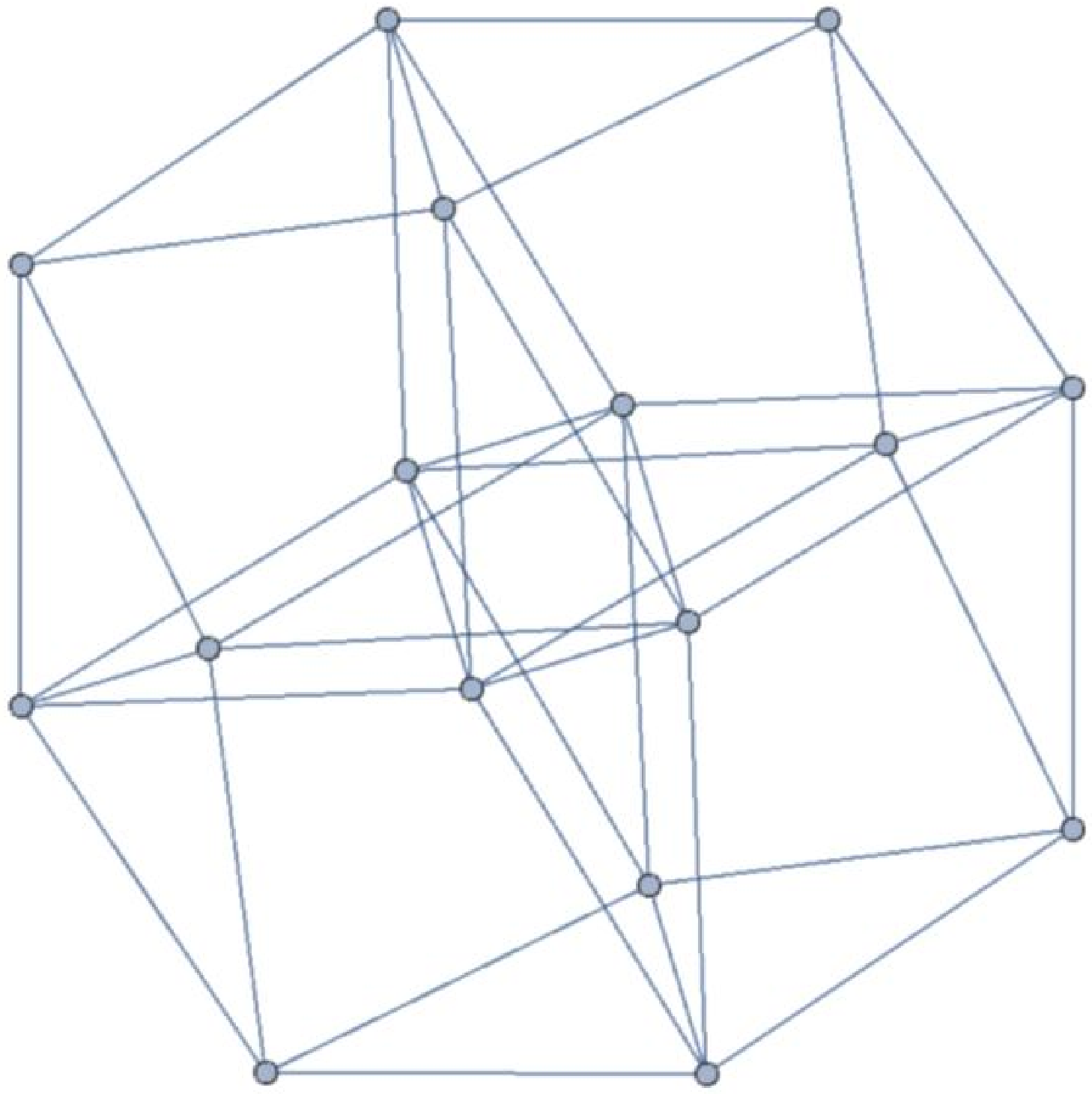}(c)\includegraphics[scale=0.22]{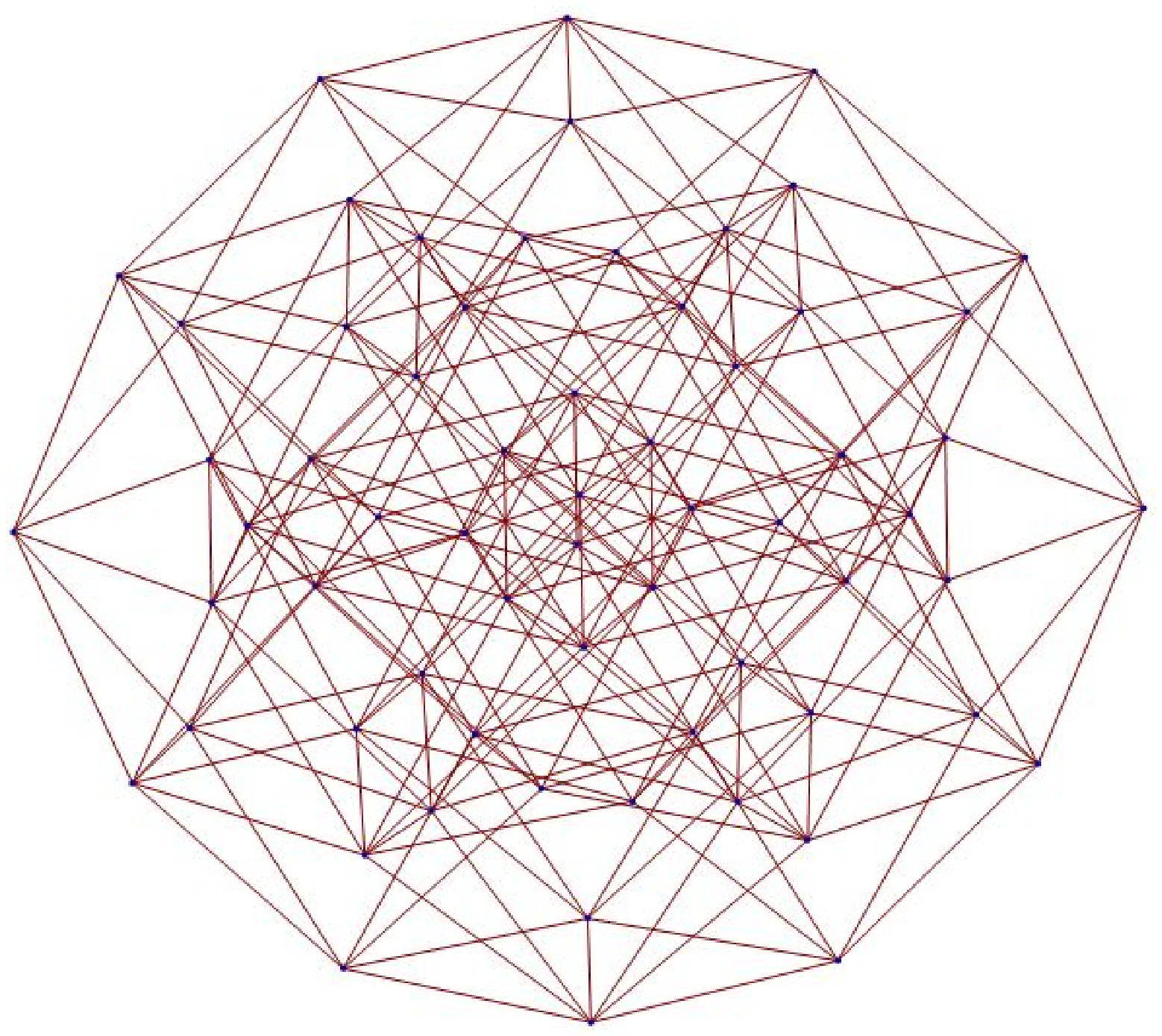}
\par\end{center}%
\end{minipage}
\par\end{centering}

\caption{\label{graph1}The graph representation of stochastic pumps with $4$
sites and $5$ bidirectional transitions: (a) the single particle
pump, (b) and (c) the product graph representing a system with $2$ and $3$ particles respectively. The nodes
of the product graphs correspond to many particle states such as $(\alpha,\gamma,\alpha)$. }
\end{figure}

It is often helpful to represent the jump process using a graph. The
nodes of this graph correspond to the coarse-grained states, also referred to as
sites. The links represent the possible (bidirectional) transitions
between sites. An example of a graph representing a $4$ site system
with $5$ possible transitions is depicted in Fig.~\ref{graph1}a.

\noindent The NPT was derived for thermally activated transitions
with rates  $R_{\beta\alpha}=k\exp\left[\left(E_{\alpha}-B_{\alpha\beta}\right)/T\right]$,
where $E_{\alpha}$ are site energies and $B_{\alpha\beta}$
are barrier energies, while $T$ is the temperature of the environment.
This parametrization of rates naturally describes certain molecular machines \cite{Kay2007,Leigh2003}, but can be used for any system satisfying
detailed balance. When the set of parameters $\left\{ E_{\alpha},B_{\alpha\beta}\right\} $
do not vary with time the system relaxes to an equilibrium distribution
with $p_{\alpha}^{(eq)}\propto e^{-E_{\alpha}/T}$. Note that detailed balance,
$R_{\beta\alpha} p_{\alpha}^{(eq)}=R_{\alpha\beta} p_{\beta}^{(eq)}$, requires symmetric barriers,
$B_{\alpha\beta}=B_{\beta\alpha}$.

When the system is driven by a time periodic variation of the site
and barrier energies, with period $\tau$, Floquet theory states
that the system settles into a time periodic
asymptotic state \cite{Talkner1999}. In what follows we always assume
that the system is in this asymptotic state.

During a cycle of the driving the probability can slosh back and forth
between sites. As a result, the existence or lack thereof of net directed
motion is determined by the integrated fluxes
\begin{equation}
\phi_{\alpha\beta}\equiv\oint J_{\alpha\beta}(t)dt,
\end{equation}
where $J_{\alpha\beta}(t)=R_{\alpha\beta}(t)p_{\beta}(t)-R_{\beta\alpha}(t)p_{\alpha}(t)$
is the momentary probability current flowing from $\beta$ to $\alpha$.
The integral over time is taken over a full period of the driving
cycle. A non vanishing $\phi_{\alpha\beta}$ for any pair of $\alpha,\beta$
means that the variation of parameters have resulted in some directed
motion. For the catenanes of \cite{Leigh2003} the directed motion
corresponds to rotation of the smaller ring like molecule along the
larger molecule. More generally this directed motion will express
the existence of some physical effect which can be utilised.

The NPT \cite{Rahav2008} states that one needs to vary {\em both}
the barrier and site energies to generate directed motion. In particular,
when one varies only $\left\{ E_{\alpha}(t)\right\} $, while the
$\left\{ B_{\alpha\beta}\right\} $ are kept fixed, no directed motion
is generated. This is non intuitive, since it seems natural that one
should be able to induce directed motion by designing a driving cycle
in which the most energetically favourable site changes in a cyclic
way, for example between sites $\alpha\rightarrow\beta\rightarrow\delta\rightarrow\alpha$
in Fig. \ref{graph1}a.

\paragraph*{Many particle stochastic pumps-}
Consider a set of $N$ distinguishable particles which make stochastic
jumps between a set of sites. The state of this many particle system
is described by specifying the location of each particle, $\bm{X}\equiv(x_{1},x_{2},\cdots,x_{N})$,
where $x_{i}$ denotes the site in which the $i$th particle resides.
For example, in the $4$ site system depicted in Fig. \ref{graph1}a,
$\bm{X}\equiv(\gamma,\delta)$ describes a state in which particle
$1$ $(2)$ is in site $\gamma$ $(\delta)$ respectively.

Transitions between many particle states are made via jumps of a single
particle. Instantaneous jumps of several
particles are neglected. We therefore adopt a notation that matches
the nature of transitions. Let $\bm{X}_{\left.\right|k}\equiv(x_{1},x_{2},\cdots,x_{k-1},x_{k+1}\cdots x_{N})$
denote the location of all particles except the $k$th one. Then $\bm{X}_{\left.\right|k};x_{k}\rightarrow x_{k}^{\prime}$
conveniently labels the transition out of $\bm{X}$ in which the $k$th
particle jumps to state $x_{k}^{\prime}$.

The fact that only single particle jumps connect the many particle
states means that the graph representing the many particle stochastic
pump is constructed by a Cartesian product of single particle graphs
\cite{graphbook}. Fig. \ref{graph1}(b) and (c) depicts the many particle
(or product) graphs corresponding to two or three particles residing
in the sites of Fig. \ref{graph1}a. It is clear that when $N\gg1$
this product graph becomes quite complicated.

Which transition rates should one expects in this type of many particle
system? For non interacting particles the total energy of the system is
\begin{equation}
\mathcal{E}(\bm{X})=\sum_{\sigma}n_{\sigma}(\bm{X})E_{\sigma},\label{eq:manybenergyni}
\end{equation}
where the summation is over the sites, and $n_{\sigma}(\bm{X})$ is
the number of particles residing in site $\sigma$. For any transition we can also assign a many particle energy
barrier. The barrier for the transition $\bm{X}_{\left.\right|k};\alpha\rightarrow\beta$
is
\begin{equation}
\mathcal{B}(\bm{X}^{\prime},\bm{X})
=\mathcal{E}(\bm{X}_{\left.\right|k})+B_{\beta\alpha}.
\end{equation}
The non-vanishing transition rates are then given by
\begin{equation}
\mathcal{R}_{\bm{X}^{\prime},\bm{X}}=\mathcal{R}_{\bm{X}_{\left.\right|k};x_{k}\rightarrow x_{k}^{\prime}}=k\exp\left[\left(\mathcal{E}(\bm{X})-\mathcal{B}(\bm{X}^{\prime},\bm{X})\right)/T\right].\label{eq:mbrates}
\end{equation}
It is easy to verify that in absence of interactions this rate is
given by $k\exp\left[(E_{\alpha}-B_{\beta\alpha})/T\right]$, just
like a single particle pump. The resulting dynamics of a pump with
$N$ non-interacting particles is that of $N$ copies
of a single particle pumps, see Sec. I of the supplementary material.

Interactions will modify the expressions for the many particle energies
and barriers. We consider short range, local, interactions, so that
a particle is only affected by particles residing at the same site.
Within the same spirit, a particle which 'crosses the barrier' between
sites does not interact with other particles. The transition rates are then given by (\ref{eq:mbrates}) with
\begin{eqnarray}
\mathcal{E}(\bm{X}) & = & \sum_{\sigma}n_{\sigma}(\bm{X})E_{\sigma}+U_{\sigma}(n_{\sigma}(\bm{X})),\\
\mathcal{B}(\bm{X}^{\prime},\bm{X}) & = & \sum_{\sigma}n_{\sigma}(\bm{X}_{\left.\right|k})E_{\sigma}+U_{\sigma}(n_{\sigma}(\bm{X}_{\left.\right|k}))+B_{\beta\alpha},
\end{eqnarray}
where $\mathcal{B}(\bm{X}^{\prime},\bm{X})=\mathcal{B}(\bm{X},\bm{X}^{\prime})$.
We note that $\mathcal{E}(\bm{X})-\mathcal{B}(\bm{X}^{\prime},\bm{X})=E_{\alpha}-B_{\beta\alpha}+\tilde{U}_{\alpha}(n_{\alpha})$
with $\tilde{U}_{\alpha}(n_{\alpha})\equiv U_{\alpha}(n_{\alpha})-U_{\alpha}(n_{\alpha}-1)$,
recasting the rates in terms of single site quantities. When $\{E_{\alpha}\},\{U_{\alpha}\}$
do not vary in time the system relaxes to an equilibrium distribution
with $p^{(eq)}(\bm{X})\propto e^{-\mathcal{E}(\bm{X})/T}$.

Since the energy of a single particle in a site is already accounted
for, we must take $U_{\alpha}(0)=U_{\alpha}(1)=0$. Otherwise, we
allow the interaction to take any value, implicitly assuming that
it is not strong enough to destabilize or create sites.

\paragraph*{Many particle fluxes and particle currents-}
We first examine the immediate application of the NPT to this many
particle pump. The pump is driven by periodic variation of $\left\{ E_{\alpha}(t)\right\} $
and $\left\{ B_{\alpha\beta}(t)\right\} $. The NPT of \cite{Rahav2008}
states that no directed motion is generated if either all many particle
energies $\mathcal{E}(\bm{X})$, or alternatively all the $\mathcal{B}(\bm{X}^{\prime},\bm{X})$
are time independent. We emphasize that this result is not very useful,
since i) the expression for $\mathcal{B}(\bm{X}^{\prime},\bm{X})$
mixes site energies and barriers, and ii) the integrated flux $\phi_{\bm{X}_{\left.\right|k};x_{k}\rightarrow x_{k}^{\prime}}$
is not the observable one typically wishes to study, since it has
the interpretation of the net probability probability flux that particle
$k$ jumps from $x_{k}$ to $x_{k}^{\prime}$ {\em conditioned on}
$\bm{X}_{\left.\right|k}$.

It is much more natural to use the net current of particles flowing
between two given sites as a measure of directed motion. This current
is simply the sum of all many particle fluxes corresponding to a given
transition
%\begin{equation}
%J_{\sigma\rightarrow\sigma^{\prime}}\equiv\sum_{k=1}^{N}\sum_{\bm{X}_{\left.\right|k}}J_{\bm{X}_{\left.\right|k};\sigma\rightarrow\sigma^{\prime}}.
%\end{equation}
%Integration over a full period of the driving gives a similar relation
%for time integrated currents
\begin{equation}
\Phi_{\sigma\rightarrow\sigma^{\prime}}=\sum_{k=1}^{N}\sum_{\bm{X}_{\left.\right|k}}\phi_{\bm{X}_{\left.\right|k};\sigma\rightarrow\sigma^{\prime}}.\label{eq:netparticlec}
\end{equation}
The main result of this letter is that the integrated particle currents
(\ref{eq:netparticlec}) satisfy {\em an additional NPT} which
is expressed in terms of the (single particle) site energies and barriers.
It states that both $\left\{ E_{\alpha}\right\} $ and $\left\{ B_{\alpha\beta}\right\} $
need to be periodically varied in time in order to generate directed
particle currents ($\Phi_{\sigma\rightarrow\sigma^{\prime}}\neq0$).

\paragraph*{Derivation of NPT-}

When the $\left\{ E_{\alpha}\right\}$ are time independent the system
relaxes to an equilibrium state with no currents. We therefore focus on
the case of time independent barriers and time dependent energies.
One of the derivations of the single particle NPT, due to Mandal and
Jarzynski \cite{Mandal2011}, is based on incompatibility of two sets
of equations, termed conservation laws and cycle equations. This graph
based approach, which we employ here, will highlight the difference
between the single particle graph, such as the one in Fig. \ref{graph1}a,
and the many particle product graph. The derivation below is based
on equations which hold on the much simpler one particle graph.

The first set of equations expresses conservation laws. For the many
particle system the master equation can be rewritten as $\frac{dP(\bm{X})}{dt}=-\sum_{\bm{X}^{\prime}\neq\bm{X}}J_{\bm{X}^{\prime},\bm{X}}$.
The periodicity of $P(\bm{X},t)$ can be used to
obtain an equation $0=\sum_{\bm{X}^{\prime}\neq\bm{X}}\phi_{\bm{X}^{\prime},\bm{X}}$
for any state $\bm{X}$. To derive equations which constrain the particle
currents we examine the density of particles at the different sites
\begin{equation}
\rho_{\sigma}\equiv\sum_{k=1}^{N}\sum_{\bm{X}_{\left.\right|k}}P(x_{1},x_{2},\cdots,x_{k-1},\sigma,x_{k+1}\cdots x_{N}).\label{eq:defdensity}
\end{equation}
We note that periodicity of $P$ implies periodicity of $\rho$. This
leads to the conservation laws,
\begin{equation}
\sum_{\sigma^{\prime}\neq\sigma}\Phi_{\sigma\rightarrow\sigma^{\prime}}=0,\label{eq:cons}
\end{equation}
expressing the fact that $\rho$ changes through
particle currents. Eq. (\ref{eq:cons}) is what one would naively guess, but it can also be derived from the conservation
laws for many particle probabilities by explicitly performing the sum
(\ref{eq:defdensity}), see Sec. II of the supplementary
information.

The second set of equations are cycle equations, defined on closed
cycles of transitions on the graph representing the system. Let $\bm{X}^{\prime\prime}\rightarrow\bm{X}\rightarrow\bm{X}^{\prime}$
denote two consecutive transitions, such that in the first the $i$'th
particle jumps from $x_{i}^{\prime\prime}$ to $x_{i}$, while in
the second the $j$'th particle makes the $x_{j}\rightarrow x_{j}^{\prime}$
jump. We wish to examine the expression $e^{B_{x_{i}^{\prime\prime},x_{i}}/T}J_{\bm{X},\bm{X}^{\prime\prime}}+e^{B_{x_{j}^{\prime},x_{j}}/T}J_{\bm{X}^{\prime},\bm{X}}$,
where we crucially use the {\em single particle barriers}. Our
goal is to see when a sum of this form over a closed cycle
will vanish. We focus on the two terms proportional to $P(\bm{X})$,
given by, $-e^{B_{x_{i}^{\prime\prime},x_{i}}/T}R_{\bm{X}^{\prime\prime},\bm{X}}P(\bm{X})+e^{B_{x_{j}^{\prime},x_{j}}/T}R_{\bm{X}^{\prime},\bm{X}}P(\bm{X})=\left[h(x_{j})-h(x_{i})\right]P(\bm{X}),$
with $h(\alpha)\equiv\exp\left[\left(E_{\alpha}+\tilde{U}_{\alpha}(n_{\alpha})\right)/T\right]$. This combination
does not depend on the barriers. This is a result of the symmetry of the barriers, which follows from detailed balance~\cite{Mandal2011}.
It is clear that the terms cancel when $x_{i}=x_{j}$. Notably, they
need not cancel when consecutive transition involve particle entering
one site and then another particle leaving a different site. As a
result not all the cycles on the product graph have cycle equations
with single particle barriers as factors.

We now focus on a particular type of cycles, in which a given particle
makes the transitions while the rest are essentially spectators. Lets
assume that the $k$'th particle makes the transitions $\alpha_{1}\rightarrow\alpha_{2}\rightarrow\cdots\alpha_{L}\rightarrow\alpha_{1}$,
closing a cycle on the corresponding single particle graph. In this
case the sum discussed above has canceling
terms, and
\begin{equation}
\sum_{i=1}^{L}e^{B_{\alpha_{i+1},\alpha_{i}}/T}J_{\bm{X}_{\left.\right|k};\alpha_{i}\rightarrow\alpha_{i+1}}=0\label{eq:intialcycle}
\end{equation}
for any $k,\bm{X}_{\left.\right|k}$. (Here $\alpha_{L+1}=\alpha_{1}$.)
This equation holds even when the barriers vary with time.

For time independent barriers an integration of (\ref{eq:intialcycle})
over a period gives an equation for integrated fluxes. Moreover, since
 the same barriers appear for different $k$'s, summation over $k,\bm{X}_{\left.\right|k}$
gives
\begin{equation}
\sum_{i=1}^{L}e^{B_{\alpha_{i+1},\alpha_{i}}/T}\Phi_{\alpha_{i}\rightarrow\alpha_{i+1}}=0.\label{eq:cycle}
\end{equation}
An equation of this type is valid for any closed cycle of transitions
on the {\em single particle graph}.

We have shown that the integrated particle currents satisfy two sets
of equations, the conservation laws (\ref{eq:cons}), and the cycle
equations (\ref{eq:cycle}). Both are defined using the nodes and
links of the single particle graph (e.g. Fig.~\ref{graph1}a)
regardless of the number of particles. The argument of \cite{Mandal2011}, which
was developed for single particle stochastic pumps, can be applied
to Eqs. (\ref{eq:cons}) and (\ref{eq:cycle}). The two sets of
equations are incompatible and the only solution is $\Phi_{\alpha\rightarrow\beta}=0$.
We conclude that the $\Phi_{\alpha\rightarrow\beta}$ satisfy an NPT: One
needs to vary both the {\em single particle} barriers, $\left\{ B_{\beta\alpha}\right\} $,
and site energies, $\left\{ E_{\alpha}\right\} $, to generated non
vanishing integrated particle currents. The interaction
clearly plays a role similar to the site energies and therefore one
could conceivably generate directed particle currents by varying the
interaction and barriers in time.

\paragraph*{Illustrative example-}

\begin{figure}
\begin{centering}
\includegraphics[scale=0.55]{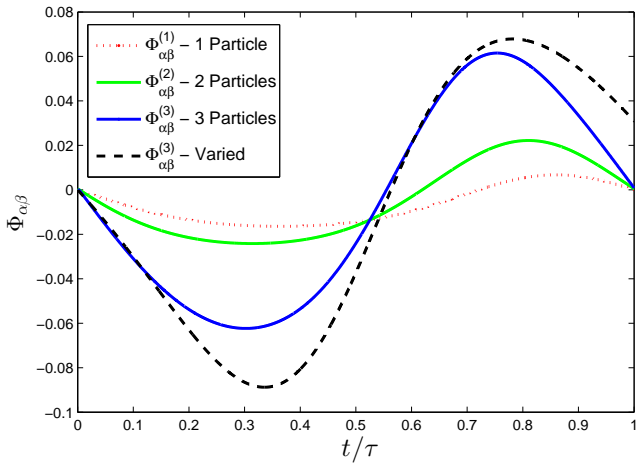}
\par\end{centering}

\caption{Time integrated particle currents for 1,2 and 3 particles between
sites $\beta\rightarrow\alpha$. Solid and dotted lines correspond to driving
cycles with time independent barriers. \label{fig:Integrated-Currents}The
dashed line depicts the integrated current when one of the barriers
is also varied in time.}
\end{figure}

To illustrate our results we have investigated the system represented
by Fig. \ref{graph1} numerically. The system was driven using
\begin{equation}
E_{i}\left(t\right)=-2+\cos\left(\frac{2\pi}{\tau}\left(t+\frac{i-1}{4}\right)\right),
\end{equation}
where $i=1$ correspond to $\alpha$, $2$ correspond to $\beta$,
etc. The energy barriers were $B_{\alpha\beta}=-0.3,\, B_{\beta\gamma}=0,\, B_{\gamma\alpha}=1 ,\, B_{\alpha\delta}=0.2$,
and $B_{\gamma\delta}=-0.1$. When both energies and barriers were
varied in time (dashed line in Fig. \ref{fig:Integrated-Currents})
$B_{\alpha\beta}\left(t\right)=\cos\left(\frac{2\pi}{\tau}\left(t+0.11\right)\right)$,
while the rest of the barriers were unchanged.
The interaction were taken to be $U(2)=\ln 2$
 and $U(3)=\ln 3$.

\begin{figure}
\begin{centering}
\includegraphics[scale=0.58]{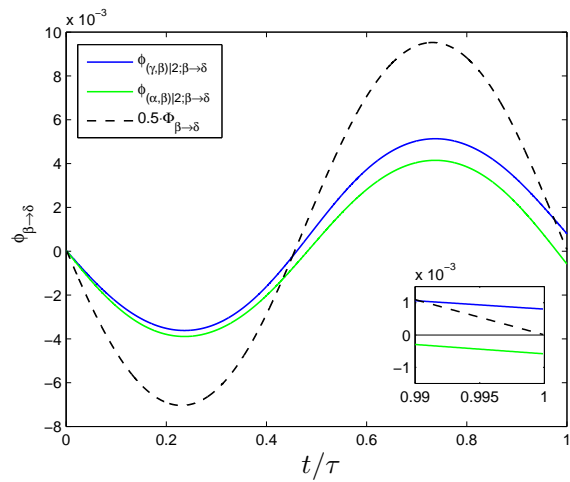}
\par\end{centering}

\caption{\label{fig:family_of_interstate_currents} Integrated fluxes corresponding
to the $\beta\rightarrow\delta$ transitions in the two particle pump of $\ref{graph1}$(b).
Solid lines correspond to many particles fluxes. The dashed line depicts the integrated particle flux $\Phi_{\beta\rightarrow\delta}$
divided by $2$. The inset shows the region near $t=\tau$.}
\end{figure}

Fig. \ref{fig:Integrated-Currents} compares the integrated particle
currents, $\Phi_{\alpha\beta} (t) \equiv \int_0^t J_{\alpha\beta} (t^\prime) dt^\prime$, for a pump with $1$, $2$ and $3$ particles. For time independent barriers
 the integrated particle currents vanishes after a full
period, as predicted by the NPT. (For one particle this is the
NPT of \cite{Rahav2008}.) The dashed curve demonstrates that directed
motion is generated when both barriers and site energies are varied.

Fig. \ref{fig:family_of_interstate_currents} depicts the integrated
many particle probability fluxes corresponding to the $\beta\rightarrow\delta$
transition (fixed barriers). It is clear the $\phi$'s
 need not vanish after full period, even when the NPT holds for the net particle currents $\Phi_{\sigma \rightarrow \sigma^\prime}$ (such as the dashed line in Fig. \ref{fig:family_of_interstate_currents}). This expresses the fact
that not all cycles on the {\em product graph} have cycle equations.

\paragraph*{Summary-}

We have studied stochastic pumps with {\em several interacting particles} and found that
the net particle current satisfies an NPT. We emphasize
 the NPT of \cite{Rahav2008} would give a condition on variation
 of many particle parameters $\left\{ \mathcal{ E,B} \right\}$ while here
 the condition uses the more natural site parameters $\left\{  E,B \right\}$.

Interactions in stochastic pumps are rarely considered. In \cite{Chernyak2011}
an open system with interacting particles was studied, and the large deviation
function was shown to be equivalent to that of a non interacting system when branching ratios were time independent.
The two papers deal with different setups.
Our stochastic pumps conserve the number of particles and
the NPT is not asymptotic in time. Nevertheless, in both systems
a simplification was achieved for time-independent branching ratios. (Here for particles
leaving the {\em same} site.) This hints that there is an underlying structure in such interacting systems
which merits further study.

%Several experimental approaches may be used to realize the dynamics of interacting stochastic particle discussed above. The experiment of Ref.~\cite{Leigh2003} included interactions which two smaller ring-like molecules were synthesized on the larger molecule. However, in that experiment the rotation of the molecules was calculated using a model rather than observed and the interaction was assumed to be exclusion.

Weakly coupled triple quantum dots \cite{Hsieh2012} offer a possible experimental realization of 
interacting particles obeying stochastic dynamics, especially when the temperature is larger than the mean level spacing.
The system can be controlled by gate voltages, although the transition rates may differ from those considered here. 
 A system of several colloidal particles in water offer another natural realization.
  Holographic techniques allow to create a controllable potential landscape \cite{Polin2005,Yevnin2013} for the colloids, including one composed of several adjacent traps. Moreover, particles in the same trap interact sterically.

\paragraph*{Acknowledgements}

We are grateful for support from the US-Israel Binational Science
Foundation (grant 2010363), and the Israel Science Foundation
(grant 924/11).


\begin{thebibliography}{27}
\expandafter\ifx\csname natexlab\endcsname\relax\def\natexlab#1{#1}\fi
\expandafter\ifx\csname bibnamefont\endcsname\relax
  \def\bibnamefont#1{#1}\fi
\expandafter\ifx\csname bibfnamefont\endcsname\relax
  \def\bibfnamefont#1{#1}\fi
\expandafter\ifx\csname citenamefont\endcsname\relax
  \def\citenamefont#1{#1}\fi
\expandafter\ifx\csname url\endcsname\relax
  \def\url#1{\texttt{#1}}\fi
\expandafter\ifx\csname urlprefix\endcsname\relax\def\urlprefix{URL }\fi
\providecommand{\bibinfo}[2]{#2}
\providecommand{\eprint}[2][]{\url{#2}}

\bibitem[{\citenamefont{Howard}(2001)}]{Howardbook}
\bibinfo{author}{\bibfnamefont{J.}~\bibnamefont{Howard}},
  \emph{\bibinfo{title}{Mechanics of Motor Proteins and the Cytoskeleton}}
  (\bibinfo{publisher}{Sinauer Associates}, \bibinfo{address}{Sunderland,
  Massachusetts}, \bibinfo{year}{2001}).

\bibitem[{\citenamefont{Kottas et~al.}(2005)\citenamefont{Kottas, Clarke,
  Horinek, and Michl}}]{Kottas2005}
\bibinfo{author}{\bibfnamefont{G.~S.} \bibnamefont{Kottas}},
  \bibinfo{author}{\bibfnamefont{L.~I.} \bibnamefont{Clarke}},
  \bibinfo{author}{\bibfnamefont{D.}~\bibnamefont{Horinek}}, \bibnamefont{and}
  \bibinfo{author}{\bibfnamefont{J.}~\bibnamefont{Michl}},
  \bibinfo{journal}{Chemical Reviews} \textbf{\bibinfo{volume}{105}},
  \bibinfo{pages}{1281} (\bibinfo{year}{2005}).

\bibitem[{\citenamefont{Kay et~al.}(2007)\citenamefont{Kay, Leigh, and
  Zerbetto}}]{Kay2007}
\bibinfo{author}{\bibfnamefont{E.}~\bibnamefont{Kay}},
  \bibinfo{author}{\bibfnamefont{D.}~\bibnamefont{Leigh}}, \bibnamefont{and}
  \bibinfo{author}{\bibfnamefont{F.}~\bibnamefont{Zerbetto}},
  \bibinfo{journal}{Angewandte Chemie International Edition}
  \textbf{\bibinfo{volume}{46}}, \bibinfo{pages}{72} (\bibinfo{year}{2007}).

\bibitem[{\citenamefont{Feringa}(2007)}]{Feringa2007}
\bibinfo{author}{\bibfnamefont{B.~L.} \bibnamefont{Feringa}},
  \bibinfo{journal}{The Journal of Organic Chemistry}
  \textbf{\bibinfo{volume}{72}}, \bibinfo{pages}{6635} (\bibinfo{year}{2007}).

\bibitem[{\citenamefont{Michl and Sykes}(2009)}]{Michl2009}
\bibinfo{author}{\bibfnamefont{J.}~\bibnamefont{Michl}} \bibnamefont{and}
  \bibinfo{author}{\bibfnamefont{E.~C.~H.} \bibnamefont{Sykes}},
  \bibinfo{journal}{ACS Nano} \textbf{\bibinfo{volume}{3}},
  \bibinfo{pages}{1042} (\bibinfo{year}{2009}).

\bibitem[{\citenamefont{Coskun et~al.}(2012)\citenamefont{Coskun, Banaszak,
  Astumian, Stoddart, and Grzybowski}}]{Coskun2012}
\bibinfo{author}{\bibfnamefont{A.}~\bibnamefont{Coskun}},
  \bibinfo{author}{\bibfnamefont{M.}~\bibnamefont{Banaszak}},
  \bibinfo{author}{\bibfnamefont{R.~D.} \bibnamefont{Astumian}},
  \bibinfo{author}{\bibfnamefont{J.~F.} \bibnamefont{Stoddart}},
  \bibnamefont{and} \bibinfo{author}{\bibfnamefont{B.~A.}
  \bibnamefont{Grzybowski}}, \bibinfo{journal}{Chem. Soc. Rev.}
  \textbf{\bibinfo{volume}{41}}, \bibinfo{pages}{19} (\bibinfo{year}{2012}).

\bibitem[{\citenamefont{Reimann}(2002)}]{Reimann2002}
\bibinfo{author}{\bibfnamefont{P.}~\bibnamefont{Reimann}},
  \bibinfo{journal}{Physics Reports} \textbf{\bibinfo{volume}{361}},
  \bibinfo{pages}{57 } (\bibinfo{year}{2002}).

\bibitem[{\citenamefont{Parrondo}(1998)}]{Parrondo1998}
\bibinfo{author}{\bibfnamefont{J.~M.~R.} \bibnamefont{Parrondo}},
  \bibinfo{journal}{Phys. Rev. E} \textbf{\bibinfo{volume}{57}},
  \bibinfo{pages}{7297} (\bibinfo{year}{1998}).

\bibitem[{\citenamefont{Sokolov}(1999)}]{Sokolov1999}
\bibinfo{author}{\bibfnamefont{I.~M.} \bibnamefont{Sokolov}},
  \bibinfo{journal}{Journal of Physics A: Mathematical and General}
  \textbf{\bibinfo{volume}{32}}, \bibinfo{pages}{2541} (\bibinfo{year}{1999}).

\bibitem[{\citenamefont{Astumian}(2003)}]{Astumian2003}
\bibinfo{author}{\bibfnamefont{R.~D.} \bibnamefont{Astumian}},
  \bibinfo{journal}{Phys. Rev. Lett.} \textbf{\bibinfo{volume}{91}},
  \bibinfo{pages}{118102} (\bibinfo{year}{2003}).

\bibitem[{\citenamefont{Sinitsyn and Nemenman}(2007)}]{Sinitsyn2007}
\bibinfo{author}{\bibfnamefont{N.~A.} \bibnamefont{Sinitsyn}} \bibnamefont{and}
  \bibinfo{author}{\bibfnamefont{I.}~\bibnamefont{Nemenman}},
  \bibinfo{journal}{Phys. Rev. Lett.} \textbf{\bibinfo{volume}{99}},
  \bibinfo{pages}{220408} (\bibinfo{year}{2007}).

\bibitem[{\citenamefont{Rahav}(2011)}]{Rahav2011}
\bibinfo{author}{\bibfnamefont{S.}~\bibnamefont{Rahav}},
  \bibinfo{journal}{Journal of Statistical Mechanics: Theory and Experiment}
  \textbf{\bibinfo{volume}{2011}}, \bibinfo{pages}{P09020}
  (\bibinfo{year}{2011}).

\bibitem[{\citenamefont{Chernyak et~al.}(2012)\citenamefont{Chernyak, Klein,
  and Sinitsyn}}]{Chernyak2012}
\bibinfo{author}{\bibfnamefont{V.~Y.} \bibnamefont{Chernyak}},
  \bibinfo{author}{\bibfnamefont{J.~R.} \bibnamefont{Klein}}, \bibnamefont{and}
  \bibinfo{author}{\bibfnamefont{N.~A.} \bibnamefont{Sinitsyn}},
  \bibinfo{journal}{The Journal of Chemical Physics}
  \textbf{\bibinfo{volume}{136}}, \bibinfo{eid}{154107} (\bibinfo{year}{2012}).

\bibitem[{\citenamefont{Sinitsyn}(2009)}]{Sinitsyn2009}
\bibinfo{author}{\bibfnamefont{N.~A.} \bibnamefont{Sinitsyn}},
  \bibinfo{journal}{Journal of Physics A: Mathematical and Theoretical}
  \textbf{\bibinfo{volume}{42}}, \bibinfo{pages}{193001}
  (\bibinfo{year}{2009}).

\bibitem[{\citenamefont{Astumian}(2011)}]{Astumian2011}
\bibinfo{author}{\bibfnamefont{R.~D.} \bibnamefont{Astumian}},
  \bibinfo{journal}{Annual Review of Biophysics} \textbf{\bibinfo{volume}{40}},
  \bibinfo{pages}{289} (\bibinfo{year}{2011}).

\bibitem[{\citenamefont{Leigh et~al.}(2003)\citenamefont{Leigh, Wong, Dehez,
  and Zerbetto}}]{Leigh2003}
\bibinfo{author}{\bibfnamefont{D.~A.} \bibnamefont{Leigh}},
  \bibinfo{author}{\bibfnamefont{J.~K.~Y.} \bibnamefont{Wong}},
  \bibinfo{author}{\bibfnamefont{F.}~\bibnamefont{Dehez}}, \bibnamefont{and}
  \bibinfo{author}{\bibfnamefont{F.}~\bibnamefont{Zerbetto}},
  \bibinfo{journal}{Nature} \textbf{\bibinfo{volume}{424}}, \bibinfo{pages}{174
  } (\bibinfo{year}{2003}).

\bibitem[{\citenamefont{Astumian}(2007)}]{Astumian2007}
\bibinfo{author}{\bibfnamefont{R.~D.} \bibnamefont{Astumian}},
  \bibinfo{journal}{Proceedings of the National Academy of Sciences}
  \textbf{\bibinfo{volume}{104}}, \bibinfo{pages}{19715}
  (\bibinfo{year}{2007}).

\bibitem[{\citenamefont{Rahav et~al.}(2008)\citenamefont{Rahav, Horowitz, and
  Jarzynski}}]{Rahav2008}
\bibinfo{author}{\bibfnamefont{S.}~\bibnamefont{Rahav}},
  \bibinfo{author}{\bibfnamefont{J.}~\bibnamefont{Horowitz}}, \bibnamefont{and}
  \bibinfo{author}{\bibfnamefont{C.}~\bibnamefont{Jarzynski}},
  \bibinfo{journal}{Phys. Rev. Lett.} \textbf{\bibinfo{volume}{101}},
  \bibinfo{pages}{140602} (\bibinfo{year}{2008}).

\bibitem[{\citenamefont{Chernyak and Sinitsyn}(2008)}]{Chernyak2008}
\bibinfo{author}{\bibfnamefont{V.~Y.} \bibnamefont{Chernyak}} \bibnamefont{and}
  \bibinfo{author}{\bibfnamefont{N.~A.} \bibnamefont{Sinitsyn}},
  \bibinfo{journal}{Phys. Rev. Lett.} \textbf{\bibinfo{volume}{101}},
  \bibinfo{pages}{160601} (\bibinfo{year}{2008}).

\bibitem[{\citenamefont{Horowitz and Jarzynski}(2009)}]{Horowitz2009}
\bibinfo{author}{\bibfnamefont{J.}~\bibnamefont{Horowitz}} \bibnamefont{and}
  \bibinfo{author}{\bibfnamefont{C.}~\bibnamefont{Jarzynski}},
  \bibinfo{journal}{Journal of Statistical Physics}
  \textbf{\bibinfo{volume}{136}}, \bibinfo{pages}{917} (\bibinfo{year}{2009}).

\bibitem[{\citenamefont{Mandal and Jarzynski}(2012)}]{Mandal2012}
\bibinfo{author}{\bibfnamefont{D.}~\bibnamefont{Mandal}} \bibnamefont{and}
  \bibinfo{author}{\bibfnamefont{C.}~\bibnamefont{Jarzynski}},
  \bibinfo{journal}{The Journal of Chemical Physics}
  \textbf{\bibinfo{volume}{137}}, \bibinfo{eid}{234104}
  (pages~\bibinfo{numpages}{7}) (\bibinfo{year}{2012}).

\bibitem[{\citenamefont{Maes et~al.}(2010)\citenamefont{Maes, NetoCny, and
  Thomas}}]{Maes2010}
\bibinfo{author}{\bibfnamefont{C.}~\bibnamefont{Maes}},
  \bibinfo{author}{\bibfnamefont{K.}~\bibnamefont{NetoCny}}, \bibnamefont{and}
  \bibinfo{author}{\bibfnamefont{S.~R.} \bibnamefont{Thomas}},
  \bibinfo{journal}{The Journal of Chemical Physics}
  \textbf{\bibinfo{volume}{132}}, \bibinfo{eid}{234116}
  (pages~\bibinfo{numpages}{6}) (\bibinfo{year}{2010}).

\bibitem[{\citenamefont{Mandal and Jarzynski}(2011)}]{Mandal2011}
\bibinfo{author}{\bibfnamefont{D.}~\bibnamefont{Mandal}} \bibnamefont{and}
  \bibinfo{author}{\bibfnamefont{C.}~\bibnamefont{Jarzynski}},
  \bibinfo{journal}{Journal of Statistical Mechanics: Theory and Experiment}
  \textbf{\bibinfo{volume}{2011}}, \bibinfo{pages}{P10006}
  (\bibinfo{year}{2011}).

\bibitem[{\citenamefont{Ren et~al.}(2011)\citenamefont{Ren, Chernyak, and
  Sinitsyn}}]{Ren2011}
\bibinfo{author}{\bibfnamefont{J.}~\bibnamefont{Ren}},
  \bibinfo{author}{\bibfnamefont{V.~Y.} \bibnamefont{Chernyak}},
  \bibnamefont{and} \bibinfo{author}{\bibfnamefont{N.~A.}
  \bibnamefont{Sinitsyn}}, \bibinfo{journal}{Journal of Statistical Mechanics:
  Theory and Experiment} \textbf{\bibinfo{volume}{2011}},
  \bibinfo{pages}{P05011} (\bibinfo{year}{2011}).

\bibitem[{\citenamefont{Talkner}(1999)}]{Talkner1999}
\bibinfo{author}{\bibfnamefont{P.}~\bibnamefont{Talkner}},
  \bibinfo{journal}{New Journal of Physics} \textbf{\bibinfo{volume}{1}},
  \bibinfo{pages}{4} (\bibinfo{year}{1999}).

\bibitem[{\citenamefont{Imrich et~al.}(2008)\citenamefont{Imrich, Klavzar, and
  Rall}}]{graphbook}
\bibinfo{author}{\bibfnamefont{W.}~\bibnamefont{Imrich}},
  \bibinfo{author}{\bibfnamefont{S.}~\bibnamefont{Klavzar}}, \bibnamefont{and}
  \bibinfo{author}{\bibfnamefont{D.~F.} \bibnamefont{Rall}},
  \emph{\bibinfo{title}{Topics in Graph Theory: Graphs and Their Cartesian
  Product}} (\bibinfo{publisher}{Taylor \& Francis Inc},
  \bibinfo{address}{Natick, USA}, \bibinfo{year}{2008}).

\bibitem[{\citenamefont{Chernyak et~al.}(2011)\citenamefont{Chernyak, Chertkov,
  and Sinitsyn}}]{Chernyak2011}
\bibinfo{author}{\bibfnamefont{V.~Y.} \bibnamefont{Chernyak}},
  \bibinfo{author}{\bibfnamefont{M.}~\bibnamefont{Chertkov}}, \bibnamefont{and}
  \bibinfo{author}{\bibfnamefont{N.~A.} \bibnamefont{Sinitsyn}},
  \bibinfo{journal}{Journal of Statistical Mechanics: Theory and Experiment}
  \textbf{\bibinfo{volume}{2011}}, \bibinfo{pages}{P09006}
  (\bibinfo{year}{2011}).

\bibitem{Hsieh2012}
C.-Y. Hsieh, Y.-P. Shim, M. Korkusinski, and P. Hawrylak, {\em Rep. Prog. Phys}, {\bf 75}, 114501 (2012).

\bibitem{Polin2005}
M. Polin, K. Ladavec, S.-H. Lee, Y. Roichman, and D. G. Grier, {\em Opt. Expr.}, {\bf 4}, 5831 (2005).

\bibitem{Yevnin2013}
M. Yevnin, D. Kasimov, Y. Gluckman, Y. Ebenstein, and Y. Roichman, {\em Biomed. Opt. Expr.}, {\bf 4}, 2087 (2013).

\end{thebibliography}
\end{document}